\documentclass[twocolumn,showpacs,amsfonts,aps,prl,floatfix,superscriptaddress]{revtex4} 
\usepackage{epsfig}
\usepackage{bm}

\newcommand{\nc}{\newcommand}

\nc{\2}{\frac{1}{2}}
\nc{\half}{{\textstyle\2}}
\nc{\3}{\frac{1}{3}}
\nc{\4}{\frac{1}{4}}

\nc\be{\begin{equation}}
\nc\ee{\end{equation}}
\nc{\beq}[1]{\begin{equation}\label{#1}}
\nc{\eeq}{\end{equation}}
\nc{\bea}[1]{\begin{eqnarray}\label{#1}}
\nc{\eea}{\end{eqnarray}}

\nc{\bce}{\begin{center}}
\nc{\ece}{\end{center}}
\nc{\bit}{\begin{itemize}}
\nc{\eit}{\end{itemize}}
\nc{\bmp}{\begin{minipage}}
\nc{\emp}{\end{minipage}}
\nc{\se}{\section}
\nc{\suse}{\subsection}

\nc{\bb}{\bm{b}}
\nc{\bq}{\bm{q}}
\nc{\bK}{\bm{K}}
\nc{\br}{\bm{r}}
\nc{\bs}{\bm{s}}

\nc{\Eq}{{\,=\,}}
\nc{\Kt}{K_\perp}
\nc{\pt}{p_\perp}
\nc{\mt}{m_\perp}
\nc{\pL}{p_{\rm L}}
\nc{\ET}{E_{\rm T}}
\nc{\Nch}{N_{\rm ch}}
\nc{\Nc}{N_{\rm coll}}
\nc{\Np}{N_{\rm part}}
\nc{\Atanh}{{\rm Atanh}}
\nc{\Asinh}{{\rm Asinh}}
\nc{\Acosh}{{\rm Acosh}}
\nc{\scm}{\sqrt{s_{\rm NN}}}

\newcommand{\gapp}{\raisebox{-.5ex}{$\stackrel{>}{\sim}$}}

\nc{\la}{\langle}
\nc{\lla}{\left \langle}
\nc{\ra}{\rangle}
\nc{\rra}{\right \rangle}

\begin{document}

\title{Emission angle dependent pion interferometry at RHIC and beyond}

\author{Ulrich Heinz}
\affiliation{Physics Department, The Ohio State University, Columbus, OH 43210}
\affiliation{Kavli Institute for Theoretical Physics, University of
             California, Santa Barbara, CA 93106-4030}
\author{Peter F. Kolb}
\affiliation{Kavli Institute for Theoretical Physics, University of
             California, Santa Barbara, CA 93106-4030}
\affiliation{Department of Physics and Astronomy, State University of 
             New York, Stony Brook, NY 11794-3800}

\begin{abstract}
We use hydrodynamics to generate freeze-out configurations for non-central
heavy-ion collisions at present and future collider energies. Such 
collisions are known to produce strong elliptic flow. The accompanying 
space-time structure of the source at freeze-out is analyzed using pion 
interferometry. Between RHIC and LHC energies the source deformation 
in the transverse plane changes sign. This leaves characteristic 
signatures in the emission angle dependence of the HBT radii.  
\end{abstract}

\vspace{1.0cm}
\pacs{25.75.-q, 25.75.Gz, 24.10.Nz} 

\date{July 10, 2002 (revised version)}

\maketitle

%
%

Heavy-ion collision data from the Relativistic Heavy Ion Collider 
(RHIC) \cite{QM01} have produced strong indications for a large degree 
of thermalization in these collisions. Only dynamical models with very 
intense reinteractions among the produced particles 
\cite{Molnar,Lin,Humanic}, in particular the hydrodynamic model 
\cite{O92,BD00,KSH00,Kolbv2,SBD01,Teaneyv2,Hirano}, have been able to 
reproduce the large observed elliptic flow signals 
\cite{STARPRL86,STARv2,PHENIXv2,PHOBOSv2} together with the measured 
systematics of the shapes of the single-particle hadron spectra 
\cite{STARspectra,PHENIXspectra} which require strong radial flow 
\cite{flow}. When compared to the data, all models seem to fail which 
are not able to achieve almost complete local thermalization very quickly,
at most 1~fm/$c$ after nuclear impact \cite{HW02,HK02}. At this early
time the energy density in the reaction zone is still far above 
1~GeV/fm$^3$ \cite{KSH00}, the critical value for color deconfinement, 
so that the thermalized state formed early in the collision has most
likely been a quark-gluon plasma \cite{HK02}.

Even though the microscopic mechanisms causing the fast thermalization 
are presently not understood, these observations provide strong 
phenomenological support for the hydrodynamical model as a dynamical 
description for the space-time evolution of the reaction zone. We will
therefore use it here to study additional aspects of the collision. We
will concentrate in particular on the question how the strong elliptic
flow affects the spatial matter distribution at ``freeze-out'' when 
the observed hadrons decouple from the fireball. Elliptic flow
requires an initial spatial deformation of the nuclear overlap region 
in the transverse plane (i.e. either a non-zero impact parameter or 
collisions between deformed nuclei) since it is generated by 
anisotropies in the pressure gradients \cite{O92,Sorge}. As it 
develops, the matter begins to expand more rapidly in the ``short'' 
direction (which for non-central collisions between spherical nuclei 
lies in the reaction plane). Hydrodynamics predicts \cite{KSH00} that, 
if the initial energy density is high enough and thus the time until 
freeze-out sufficiently long, this initial out-of-plane deformation 
changes sign, leading to an ``in-plane extended source'' (IPES) at
decoupling. We show that at RHIC energies this is not yet the case 
\cite{fn1}, but that it will eventually happen at higher collision energies 
(LHC or beyond). In this paper we discuss how this transition can be
probed by two-particle interferometry, in particular by studying the
emission angle dependence relative to the reaction plane of the size 
parameters (``HBT radii'') extracted from Bose-Einstein correlation 
functions \cite{VC95,W97,LHW00,WH99,HHLW02}. We show that the spatial 
and momentum anisotropies of the phase-space distribution at freeze-out 
(the {\em emission function} $S(x,K)$) generate characteristic 
azimuthal oscillations of the HBT radii whose phases and amplitudes 
reflect this transition. The way in which they reflect it turns out 
to be non-trivial and will be discussed in some detail. 

The hydrodynamic model and equation of state used in our calculations 
are described elsewhere \cite{KSH00}. We here study semicentral Au+Au 
collisions at impact parameter $b\Eq7$\,fm. $\bb$ is taken to point 
in $x$ direction, $z$ is the beam axis such that $(x,z)$ span the 
reaction plane, and $y$ points perpendicular to it. We assume 
boost-invariant longitudinal expansion and concentrate 
on the dynamics in the transverse $(x,y)$-plane at $z\Eq0$ where the 
nuclei collide. We use two different sets of initial conditions. The 
first set (labelled ``RHIC1'') corresponds to optimized values for 
central Au+Au collisions at $\sqrt{s}\Eq130\,A$\,GeV, which also give 
a good description of the measured hadron spectra and elliptic flow 
in non-central collisions
\cite{STARPRL86,STARv2,PHENIXv2,PHOBOSv2,STARspectra,PHENIXspectra}
out to transverse momenta of about 2\,GeV/$c$ and impact parameters
of about 10\,fm \cite{KSH00,Kolbv2,HK02}. The initial entropy density
distribution is parametrized by a superposition of two components which
scale with the number of wounded nucleons (with 75\% weight) and binary 
nucleon-nucleon collisions (with 25\% weight), respectively \cite{KHHET}.
The corresponding peak value of the initial temperature, at the starting
time $\tau_0\Eq0.6$\,fm/$c$ of the hydrodynamic expansion, in central 
($b\Eq0$) collisions is $T_0\Eq340$\,MeV, scaled down in semicentral 
collisions with the Glauber prescription described in \cite{KHHET}. 
Freeze-out is implemented with the Cooper-Frye algorithm \cite{CF74}
along a surface of constant temperature $T_{\rm f}\Eq130$\,MeV.

The second set of initial conditions (labelled ``IPES'') uses a much 
higher initial temperature $T_0\Eq2$\,GeV (at $b\Eq0$), and the 
hydrodynamic evolution is started at $\tau_0\Eq0.1$\,fm/$c$ and 
stopped when a freeze-out temperature $T_{\rm f}\Eq100$\,MeV has 
been reached. Such a high initial temperature can probably not even 
be achieved at the LHC, and the strong transverse flow generated in 
this case probably causes the system to decouple already much closer 
to the hadronization temperature of (in our case) $T_{\rm had}\Eq164$\,MeV. 
We force the system to start at such a high temperature and freeze 
out so low in order to give it enough time to convert the initial 
out-of-plane deformation into a final in-plane deformation (which is
the phenomenon we want to study). Unfortunately, over large times
and due to the strong radial flow numerical instabilities in the 
hydrodynamic code accumulate, and to keep these at a minimum we use 
(for IPES initial conditions only) an equation of state (EOS) without a 
sudden phase transition. Arguing that with these initial conditions the 
system anyway spends almost all of its time in the quark-gluon plasma 
phase, we simply use $P\Eq\3e$, corresponding to an ideal gas of 
massless quarks and gluons. After stopping the evolution at $T_{\rm f}$, 
we assume that pions are emitted with this freeze-out temperature and 
the hydrodynamic flow established on the corresponding freeze-out 
surface. This implies a change of EOS at freeze-out and is thus not 
entirely dynamically consistent; however, since modifications of the 
dynamics resulting from a proper transition to a hadron resonance gas 
EOS at hadronization are expected to have at most minor effects on 
the source anisotropy at such a late time, this dynamical 
inconsistency should not qualitatively affect our results.

\begin{figure}[t]
\epsfig{file=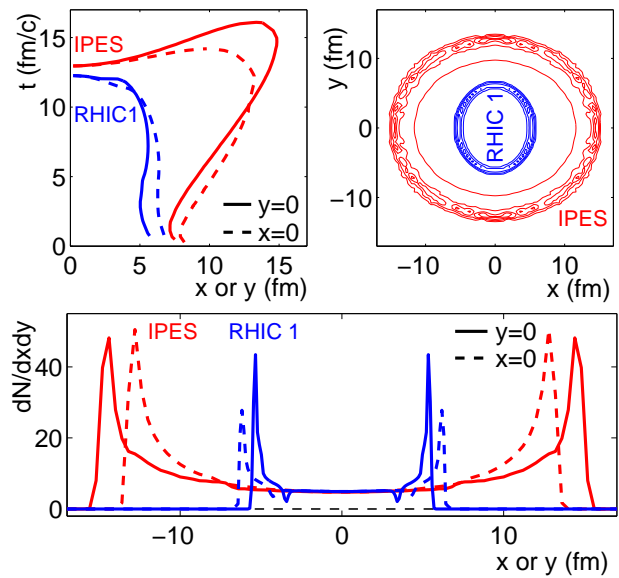,width=0.95\linewidth}
\caption{(a) Cuts through the freeze-out surface $t_{\rm f}(\br)$ 
         at $z\Eq0$ along and perpendicular to the reaction plane.
         (b) Contour plots in the transverse plane of the time-, $z$-, 
         and momentum-integrated emission function $d^2N/d^2r$ for 
         RHIC1 and IPES initial conditions (see text). 
         (c) Cuts through diagram (b) along and perpendicular to the 
         reaction plane, for RHIC1 and IPES initial conditions.
         }  
\label{F1} 
\end{figure}

The pion emission function $S(x,K)$ for pions of a given charge is 
computed from the hydrodynamic output by the Cooper-Frye prescription 
\cite{CF74}
\beq{source}
 S(x,K) =  
\frac{1}{(2\pi)^3} 
  \int_\Sigma \frac{K{\cdot}d^3\sigma(x')\,\delta^4(x{-}x')}
                   {\exp[K{\cdot}u(x')/T_{\rm f}]- 1}\,.
\nonumber
\eeq
This includes only directly emitted pions. While resonance decay pions 
strongly affect the shape of the single-particle spectrum, the same 
is not true for the HBT radii extracted from the two-pion correlator 
\cite{WH97}. They do, however, generate a non-Gaussian spatial tail 
in the emission function \cite{WH97,SBD01} which significantly increases 
its spatial widths (``homogeneity lengths'', see \cite{WH99}), thereby 
destroying \cite{LKP02} the direct correspondence between the latter and 
the measured HBT radii which exists for Gaussian sources \cite{WH99}. 
Since we want to preserve this correspondence as much as possible, 
we exclude decay pions when calculating the homogeneity lengths. We 
have checked that the remaining differences between the HBT radii 
calculated from the homogeneity lengths (see below) and those 
extracted from a Gaussian fit to the correlation function (calculated 
in the standard way \cite{CH94} by Fourier transforming the emission 
function) are small enough to not affect the qualitative features of
our results.

\begin{figure}[t]
\epsfig{file=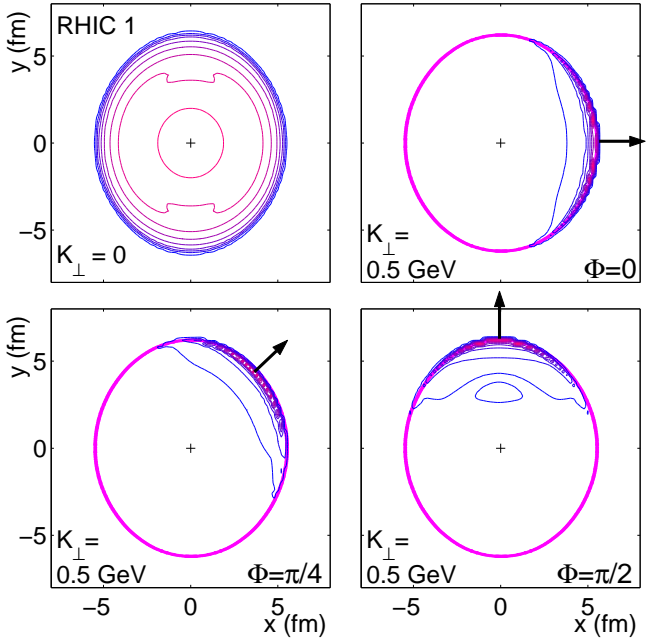,width=\linewidth}
\caption{Contours of constant emission density in the transverse plane
         for RHIC1 initial conditions (see text). The thick line 
         indicates the largest transverse extension of the freeze-out 
         hypersurface (see Figs.~\ref{F1}a,b). The four panels show 
         emission regions for midrapidity pions ($Y\Eq0$) with 
         $\Kt\Eq0$ and, for three emission angles indicated by arrows, 
         with $\Kt\Eq0.5$\,GeV.
}  
\label{F2} 
\end{figure}

In equation~(\ref{source}), $\Sigma$ denotes the freeze-out surface 
$t_{\rm f}({\bm{x}})$ of constant temperature $T_{\rm f}$, and 
$u^\mu(x')$ is the flow velocity on $\Sigma$. Figure~\ref{F1}a 
shows cuts through $\Sigma$ along ($y\Eq0$) and perpendicular to 
the reaction plane ($x\Eq0$). Initially the source is extended 
out-of-plane (larger in $y$ than in $x$ direction), but it then 
expands more rapidly into the $x$-direction, becoming in-plane 
elongated at later times. For RHIC1 initial conditions this only 
happens after most of the matter has already decoupled; as a 
consequence the time-integrated source, shown in Fig.~\ref{F1}b, 
is still longer in $y$ than in $x$ direction in the RHIC1 case. For 
IPES initial conditions the deformation changes sign before most 
particles decouple, and the time-integrated source appears 
in-plane-extended (see again Fig.~\ref{F1}b). Also, it is much 
larger due to the much higher initial energy density and longer 
lifetime. 

Figure~\ref{F1}c shows cuts along and perpendicular to the reaction 
plane through the density contour plots 
$d^2N/d^2r\Eq\int (d^3K/E_K) dz\, dt\,S(x,K)$ of Fig.~\ref{F1}b.
Pion emission is seen to be strongly surface peaked, in particular
at RHIC1 where the freeze-out radius is almost constant for a long
time. This ``opacity'' is weaker both at lower collision energies 
(where the freeze-out surface shrinks to zero continuously 
\cite{KSH99}) and at higher energies, due to larger temporal 
variations of the freeze-out radius.

\begin{figure}[t]
\epsfig{file=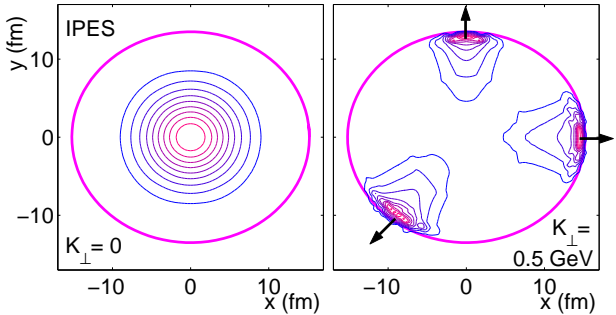,width=0.95\linewidth}
\caption{Same as Fig.~\ref{F2}, but for IPES initial conditions.
}  
\label{F3} 
\end{figure}

Figures \ref{F2} and \ref{F3} show the spatial distributions of 
pions emitted with fixed momentum. Shown are density contours of 
$\int dz\,dt\,S(x,K)$ in the transverse plane $(x,y)$ for pions
with rapidity $Y\Eq0$ and fixed $\Kt$, for three emission angles 
$\Phi\Eq0,\,45^\circ$ and $90^\circ$ relative to the reaction 
plane. (In Fig.~\ref{F3} we replaced $\Phi\Eq45^\circ$ for clarity
by the equivalent angle $\Phi\Eq225^\circ$.) Particles with vanishing 
transverse momentum $\Kt$ are seen to be emitted from almost the 
entire interior of the ``bathtub'' shown in Figs.~\ref{F1}b,c; for 
RHIC1 (IPES) initial conditions this region is elongated out-of-plane 
(in-plane). For slow pions the source thus looks transparent. Pions 
with sufficiently large transverse momenta are emitted from relatively 
thin regions close to the rim of the ``bathtub'' where the flow 
velocity is largest and points into the direction of the emitted 
pions. For fast pions the source thus looks opaque 
\cite{HV97,TH98,McLP02}. Their emission regions rotate with the 
emission angle, constrained by the shape of the ``bathtub'' even 
though they are much smaller. Again we see that the opaqueness is 
stronger at RHIC than at higher energies. It is also seen to be
anticorrelated with the curvature of the ``bathtub'' wall: where 
the curvature is large, the source is squeezed less tightly to the 
wall than where it is smaller. Fig.~\ref{F3}b shows that for IPES 
this geometric effect wins over the anisotropic flow effects which 
push the emission region more strongly towards the wall in $x$ than 
in $y$ direction; for RHIC1 both effects act together.

%
\begin{figure}[t]
\epsfig{file=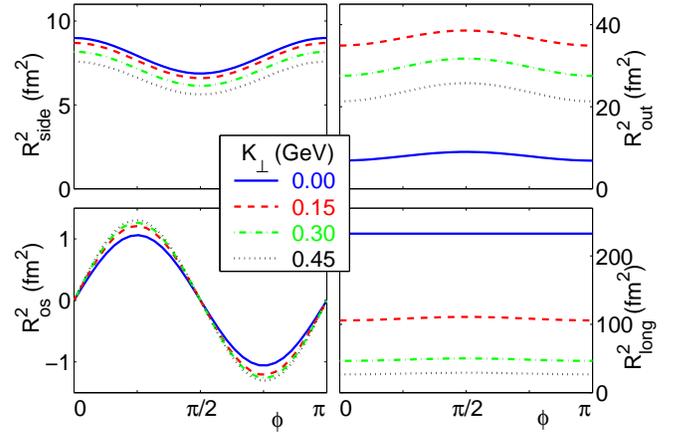,width=\linewidth}
\caption{Azimuthal oscillations of the HBT radii at $Y\Eq0$ for 
         $b\Eq7$\,fm Au+Au collisions at $\sqrt{s}\Eq130\,A$\,GeV (RHIC1),
         for four values of the transverse momentum $\Kt$ 
         as indicated.
}  
\label{F4} 
\end{figure}
%

We now proceed to compute the HBT radii from the widths of the 
$K$-dependent emission regions \cite{WH99}. The indices $o$ (for 
``outward'') and $s$ (for ``sideward'') indicate the directions 
parallel and perpendicular to the emission vector $\bm{K}_\perp$;
$l$ indicates the longitudinal $z$ direction \cite{WH99}. Given the 
spatial correlation tensor 
$S_{\mu \nu}\Eq\la x_\mu x_\nu\ra-\la x_\mu\ra\la x_\nu\ra$, 
which describes the ($K$-dependent) widths of the emission region
in space and time, the HBT radii can be calculated at midrapidity
($Y\Eq0$) from the relations \cite{W97,LHW00} 
 \bea{2}
   R_s^2 &=&\frac{S_{xx}{+}S_{yy}}{2}
           -\frac{S_{xx}{-}S_{yy}}{2}\cos(2\Phi)
           - S_{xy} \sin(2\Phi),
 \nonumber\\
   R_o^2 &=&\frac{S_{xx}{+}S_{yy}}{2}
           +\frac{S_{xx}{-}S_{yy}}{2}\cos(2\Phi)
           + S_{xy} \sin(2\Phi) 
 \nonumber\\
          && -2\beta_\perp(S_{tx}\cos\Phi{+}S_{ty}\sin\Phi) 
             +\beta_\perp^2 S_{tt},
 \nonumber\\
   R_{os}^2 &=& S_{xy}\cos(2\Phi)-\frac{S_{xx}{-}S_{yy}}{2}\sin(2\Phi)
 \nonumber\\
            &&  +\beta_\perp (S_{tx}\sin\Phi{-}S_{ty}\cos\Phi).
 \nonumber\\
   R_l^2 &=& S_{zz}.
 \eea
Note that $R_o^2$ and $R_{os}^2$ receive purely geometric 
(first lines) and mixed space-time correlation contributions 
(second lines of the equations). Since we want to use pion 
interferometry to obtain information on the geometric deformation 
of the source at freeze-out, these must be discussed separately 
(see Fig.~\ref{F5} below). 

%
\begin{figure}[t]
\epsfig{file=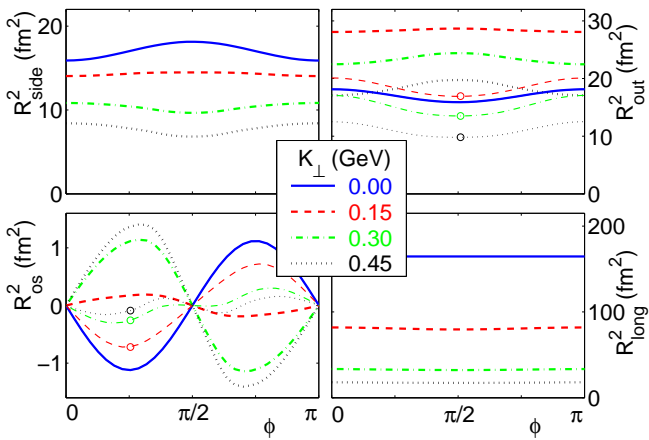,width=\linewidth}
\caption{Same as Fig.~\ref{F4}, but for IPES initial conditions.
         For $R_o^2$ and $R_{os}^2$ the geometric contributions 
         are shown separately as thin circled lines (see text).
}  
\label{F5} 
\end{figure}
%

In addition to the explicit $\Phi$ dependence exhibited in Eqs.~(\ref{2}), 
which arises from the angle $\Phi$ between the $(o,s)$ and $(x,y)$ coordinate 
systems \cite{W97}, the spatial correlation tensor $S_{\mu\nu}$ depends 
implicitly on $\Phi$ through the emission function $S(x;Y,\Kt,\Phi)$.
The combined $\Phi$ dependence of the radius parameters is shown in
Figs.~\ref{F4} and \ref{F5}, for RHIC1 and IPES initial conditions,
respectively. We first note that $R_l^2$ does not show any interesting
azimuthal oscillations and requires no further discussion. The same 
turns out to be true for the emission duration $S_{tt}$ (the last term
contributing to $R_o^2$ in Eqs.~(\ref{2})). Interesting temporal 
contributions to the oscillation patterns of $R_o^2$ and $R_{os}^2$
thus only involve the terms $S_{tx}$ and $S_{ty}$ which correlate
the freeze-out positions with time. 

Let us first discuss the geometric contributions. From Figs.~\ref{F2} 
and \ref{F3} we already know that pions with $\Kt\Eq0$ are emitted 
from almost the entire fireball and thus probe the different sign of
the spatial deformation of the {\em total} (momentum-integrated) RHIC1 
and IPES sources shown in Fig.~\ref{F1}. This is reflected by the 
opposite sign of the oscillation amplitudes of $R_s^2$ and of the 
geometric contribution to $R_o^2$ in Figs.~\ref{F4} and \ref{F5} at 
$\Kt\Eq0$. (For clarity we do not show separately the purely geometric 
contributions to $R_o^2$ and $R_{os}^2$ in Fig.~\ref{F4} because at 
RHIC1 they oscillate in the same way as the total radii, in contrast 
to the IPES case, Fig~\ref{F5}.) At higher $\Kt$-values the oscillations 
in Fig.~\ref{F5} for $R_s^2$ change sign, but those of the geometric 
contribution to $R_o^2$ do not. This reflects an intricate interplay 
between geometric and flow effects, including the already mentioned
weaker opacity for IPES.

\begin{figure}[t]
\epsfig{file=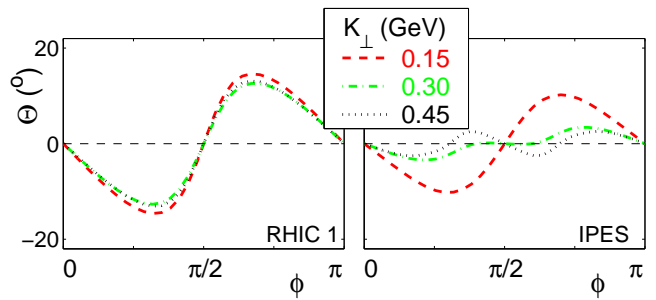,width=\linewidth}
\caption{The tilt angle $\Theta$ of the major axes of the emission region
         in the transverse plane relative to the $(o,s)$ system, for 
         different values of $\Kt$. 
}  
\label{F6} 
\end{figure}

The geometric contribution to the azimuthal oscillation of $R_{os}^2$
has an interesting intuitive interpretation: Figs.~\ref{F2} and \ref{F3}
show that for $\Kt\ne 0$ the emission region rotates together with the
emission direction $\Phi$, but not quite in phase. For each direction
$\Phi$ we can diagonalize the spatial correlation tensor in the 
transverse plane. Let us denote the major axes by $X$ and $Y$, with 
$X$ pointing approximately outward and $Y$ approximately parallel to 
the fireball surface. It is then easy to show that
 \beq{3}
  R_{os}^{2{\rm (geom)}} = \half(S_{XX}-S_{YY})\sin(2\Theta),
 \eeq
where $\Theta$ is the tilt angle between the outward-pointing major 
axis $X$ and the emission direction $\bm{K}_\perp$ (resp. the direction 
of $x_o$). Positive values of $\Theta$ correspond to counterclockwise
rotation. Figs.~\ref{F2} and \ref{F3} show that for $\Kt{\,\ne\,}0$ 
the deformation $\half(S_{XX}-S_{YY})$ of the emission region in its 
major axis frame is negative for RHIC1 but positive for IPES. For RHIC1 
$R_{os}^2$ is always positive in the first quadrant, while for IPES it
is mostly negative. According to Eq.~(\ref{3}) the tilt angle $\Theta$ is 
then negative in both cases, i.e. in the first quadrant the major axes $(X,Y)$ 
of the emission region are always rotated clockwise against the $(o,s)$ 
axes. This is confirmed by visual inspection of Figs.~\ref{F2} and \ref{F3}. 
Figure~\ref{F6} shows the tilt angle $\Theta$ as a function of the 
emission direction $\Phi$, for various values of $\Kt$. The case 
$\Kt\Eq0$ is special since the corresponding source is $\Phi$-independent. 
Then the tilt angle is exactly $\Theta\Eq{-}\Phi$ (mod $\pi$).

%
%

Last but not least we consider the contributions from the space-time
correlations $S_{tx}$ and $S_{ty}$ to the oscillation amplitudes of 
$R_o^2$ and $R_{os}^2$. They are multiplied by $\beta_\perp$ (see 
Eqs.~(\ref{2})) and thus grow with increasing $\Kt$. They reflect the 
change (caused by the elliptic flow) of the spatial source deformation 
with increasing time. Consequently their sign does not change from 
RHIC1 to IPES, contrary to the spatial contributions to the 
oscillations. For RHIC1 they reinforce the geometric oscillations of 
$R_o^2$ and $R_{os}^2$, but Fig.~\ref{F5} shows that in the IPES case 
they act against them and are responsible for changing the sign 
of the oscillations for $\Kt{\,\gapp\,}100-150$\,MeV. We interpret this
change of sign of the oscillations with increasing $\Kt$ as a 
possible experimental signature for the {\em spatial} manifestation 
of elliptic flow, i.e. the fact that the source expands faster in the 
$x$ than in the $y$ direction.

{\em Summary.} We have presented a detailed analysis of the space-time 
geometry of non-central relativistic heavy ion collisions within an ideal 
hydrodynamic framework. Although the idealized hydrodynamic picture with
Cooper-Frye freeze-out does not \cite{HK02} give a quantitative description 
of the experimental HBT radii for central Au+Au collisions at RHIC 
\cite{RHICHBT} and should eventually be replaced by a more realistic
approach to kinetic freeze-out \cite{BD00,SBD01,Teaneyv2}, we expect the 
qualitative geometric features of non-central collisions discussed here 
and their manifestation in the emission angle dependence of the HBT radii 
to be insensitive to details of how the particles decouple from the 
fireball. We have shown that at high enough collision energies the 
anisotropic flow will lead to a source which in a transverse cut is 
elongated along the reaction plane, orthogonal to the initial out-of-plane 
elongation of nuclear overlap region. As an unambiguous signature for 
this change of spatial deformation we identified a sign change of the 
oscillation amplitude as a function of emission angle of the sideward 
HBT radius $R_s$ at small transverse momentum $\Kt$. Along with this we 
point out additional changes in the oscillation patterns of $R_o$ and 
$R_{os}^2$ which are caused by a combination of geometric spatial and 
temporal aspects of the source at freeze-out. A sign change as a function 
of increasing $\Kt$ in the oscillations of $R_{os}^2$ at high collision 
energies has been interpreted as measurable evidence for the faster growth 
of the source along the reaction plane than perpendicular to it, due to 
elliptic flow. According to our calculations the effective source at RHIC 
energies should still be elongated out of the reaction plane at freeze-out. 
The corresponding predicted oscillation pattern of the HBT observables 
shown in Fig.~\ref{F4} is in qualitative agreement with preliminary 
data from RHIC \cite{R02}. Qualitatively similar conclusions as to the
sign of the source deformation at RHIC have been drawn by Lisa and 
Wells \cite{LW02}, using a hydrodynamically motivated exploding source 
model to fit the spectra \cite{STARspectra} and HBT radii 
\cite{RHICHBT,R02}.

\noindent{\em Acknowledgments:}
This work was supported in part by the U.S. Department of Energy under 
Contracts No. DE-FG02-01ER41190 and DE-FG02-88ER40388 and by the 
National Science Foundation under Grant No. PHY99-07949. We gratefully 
acknowledge the warm hospitality of the Kavli Institute for Theoretical 
Physics at the University of Santa Barbara during the Workshop Program 
``QCD in the RHIC Era''.

\bigskip



\begin{thebibliography}{99}
\bibitem{fn1}
  At late times the fireball parts which have not yet decoupled are 
  already elongated in-plane \cite{KSH00}, but the time-integrated
  source is still extended out-of-plane.
\end{thebibliography}
\end{document}